\documentclass[pre,twocolumn,showpacs,preprintnumbers,amsmath,amssymb,floatfix]{revtex4}

\usepackage[dvips]{graphicx}    
\usepackage[normalem]{ulem}       
\usepackage{natbib}
\usepackage[usenames]{color} 

\newcommand{\be}{\begin{equation}}      
\newcommand{\bea}{\begin{eqnarray}}
\newcommand{\ee}{\end{equation}}
\newcommand{\eea}{\end{eqnarray}}
\newcommand{\bdm}{\begin{displaymath}}
\newcommand{\edm}{\end{displaymath}}
\newcommand{\half}{{\textstyle \frac{1}{2}}}    

\newcommand{\ten}[1]{\uuline{#1}{}}     

\newcommand{\scp}{a_t}

\newcommand{\lda}[1]{\lambda_{\textrm{#1}}}      
\newcommand{\CLDA}[1]{\Lambda_{\textrm{#1}}}      
\renewcommand{\vec}[1]{{\mathbf #1}}        
\renewcommand{\det}[1]{{\textrm{det}}[#1]}

\begin{document}

\title{Smectic-$A$ elastomers with weak director anchoring}

\author{J. M. Adams}
\affiliation{Cavendish Laboratory, JJ Thomson Avenue, Cambridge CB3
0HE, United Kingdom}

\author{Mark Warner}
\affiliation{Cavendish Laboratory, JJ Thomson Avenue, Cambridge CB3
0HE, United Kingdom}

\author{Olaf Stenull}
\affiliation{Department of Physics and Astronomy, University of
Pennsylvania, Philadelphia, PA 19104, USA}

\author{T. C. Lubensky}
\affiliation{Department of Physics and Astronomy, University of
Pennsylvania, Philadelphia, PA 19104, USA}

\begin{abstract}

Experimentally it is possible to manipulate the director in a
(chiral) smectic-$A$ elastomer using an electric field. This
suggests that the director is not necessarily locked to the layer
normal, as described in earlier papers that extended rubber
elasticity theory to smectics. Here, we consider the case that the
director is weakly anchored to the layer normal assuming that there
is a free energy penalty associated with relative tilt between the
two. We use a recently developed weak-anchoring generalization of
rubber elastic approaches to smectic elastomers and study shearing
in the plane of the layers, stretching in the plane of the layers,
and compression and elongation parallel to the layer normal. We
calculate, inter alia, the engineering stress and the tilt angle
between director and layer normal as functions of the applied
deformation. For the latter three deformations, our results predict
the existence of an instability towards the development of shear
accompanied by smectic-$C$-like order.
\end{abstract}
\pacs{61.30.Vx, 83.80.Va, 81.40.Jj}
\maketitle


\section{Introduction}

Smectic-$A$ (Sm-$A$) liquid crystal elastomers incorporate the
anisotropic properties of liquid crystals, and the rubber elasticity
of polymer networks. The formation of a smectic layer structure by the
mesogens is the cause of their particularly anisotropic elastic
properties. These elastomers have been synthesized, and their elastic
properties explored through mechanical testing. Nishikawa and
Finkelmann found that a class of strongly coupled Sm-$A$ systems
behave like $2D$ elastomers in the layer plane \cite{nishikawa1997}
but that they are extremely stiff when stretched parallel to the layer
normal. At a threshold of a few percent strain along the layer normal,
the elastomer becomes mechanically softer, and turns opaque
\cite{nishikawa1999} because of layer rotation. This behavior is
reversible when the strain is removed. Other weakly coupled smectics
also have thresholds, but not to layer rotation, and do not have the
same extreme mechanical anisotropy \cite{Beyer:07}. A threshold has
also been reported in Sm-$A$ materials with a shorter correlation
length of the smectic layers, but here the sample remains transparent
after the threshold \cite{komp07}. Typically, smectic elastomers are
synthesized in the form of films, either with the layer normal in the
plane of the film \cite{nishikawa1997,nishikawa1999,komp07} or with
the layer normal perpendicular to the film plane \cite{nishikawa04}.
However, mechanical testing has only been performed on the first of
these two types. The second type would be useful for performing
mechanical compression tests parallel to the layer normal.

Experimental study of the electroclinic effect in Sm-$A$ elastomers
suggests that the layer normal and the director can be manipulated
with an electric field and indicates that the two are not
necessarily rigidly locked \cite{spillmann07}, at least on the scale
of electrical energies, as a Lagrangian elasticity theory developed
in Ref.~\cite{stenull:011706} assumes.
Since rubber elastic energies are typically larger than those of
electric fields, one would expect mechanical fields also
to induce relative rotations. In fact, such relative rotation has
been observed experimentally in very recent shear experiments by
Kramer and Finkelmann \cite{KraFin2008}.

On the theoretical side, a model of Sm-$A$ elastomers has been
constructed using non-linear rubber elasticity extended to smectics
\cite{adams:05a} that describes the results of Nishikawa and
Finkelmann well. This model rigidly locks the director to the layer
normal. As mentioned above, a model based on Lagrangian elasticity
theory has also been developed \cite{stenull:011706}. This model
fits well with the data, however, unlike \cite{adams:05a}, it allows
in principle for the relative rotation of the director and the layer
normal. Triggered by experiments of Kramer and Finkelmann
\cite{kramer-2007,KraFin2008} where in-plane shears were applied to
Sm-$A$ elastomers, the rubber elastic approach has been extended
\cite{stenull07} very recently to the case of soft-anchoring. In
Ref.~\cite{stenull07}, we used this model specifically to study the
shear experiments of Kramer and Finkelmann. In the present paper, we
employ this model to study various shear and stretch deformations.

The plan of presentation is the following: First, we review the
extension of non-linear rubber elasticity theory to smectics. The
fundamental distortions of imposed in-plane stretch and in-plane shear
are then explored.  Armed with these modes of deformation, we then
explore imposed extension and compression along the layer normal which
are complex but decomposable into the fundamental modes.

\section{Model Free energy}
\label{modelFreeEnergy}

The model free energy reviewed here is generalization of
  the original rubber elastic model of smectics \cite{adams:05a}, and
  was developed originally in Ref.~\cite{stenull07}. It has contributions from
the background nematic elasticity, and from the compression or
dilation of the layers. In addition to these two terms, we include
here a potential that penalizes the deviation of the director from
the layer normal \cite{PhysRevA.14.1202}:
\be
f = f_\mathrm{trace} + f_\mathrm{layer} + f_\mathrm{tilt}.
\ee
The nematic component of the free energy density has been widely
discussed \cite{warnerbook:03}, and is given by
\be
f_\mathrm{trace} = \half \mu\textrm{Tr}\left[ \ten{\lambda}\cdot
  \ten{\ell}_0 \cdot \ten{\lambda}^{T} \cdot \ten{\ell}^{-1}\right] ,
\ee
where $\mu$ is the isotropic state shear modulus, $\ten{\lambda}$ is
the deformation gradient tensor, $\ten{\ell}_0 = \ten{\delta} +
(r-1) \vec{n}_0 \vec{n}_0$ is the step-length tensor before the
deformation has been applied, $\ten{\ell}^{-1} = \ten{\delta}+
(1/r-1) \vec{n}\vec{n}$ is the inverse of the step-length tensor
after the deformation has been applied, and $\ten{\delta}$ is the
unit tensor. The step length-tensor is proportional to the second
moment of the Gaussian distribution of anisotropic chains making up
the rubbery network. The anisotropy of the chains is parameterized
by $r$. The smectic layers embedded in the elastic matrix give rise
to the following contribution to the free energy
\be
f_\mathrm{layer} = \half B \left( \frac{d}{d_0 \cos \Theta} -1
\right)^2,\label{eq:fsm}
\ee
where $B$ is the layer spacing modulus (typically larger than in
liquid smectics), $d$ is the current layer spacing, $d_0$ is the
natural layer spacing and $\Theta$ is the angle between the layer
normal and the director. For smectics where the layers are
strongly coupled to the matrix, changes in layer spacing can be
derived by analyzing how the embedded layer normal deforms in
step with the elastic matrix. The layer normal $\vec{k}$
and layer spacing are given by \cite{adams:05a}
\be
\vec{k} = \frac{\ten{\lambda}^{-T} \cdot
  \vec{k}_0}{|\ten{\lambda}^{-T} \cdot \vec{k}_0|}
\;\;\; ; \;\;\; \frac{d}{d_0} = \frac{1}{|\ten{\lambda}^{-T} \cdot
\vec{k}_0|} \label{eq:layers}\ee
where $\vec{k}_0$ is the initial layer normal. The deformation of
the layer normal outlined above affinely follows that of the rubber
matrix because the energetic penalty of not doing so scales with the
system size in the microscopic model described in \cite{adams:05a}.
The $\cos \Theta$ term reflects the projection of the rods making up
the layer spacing contracting as the director tilts. Note that this
term does not include the finite thickness of the rods, since as
$\Theta\rightarrow \pi/2$ then $d\rightarrow0$ to avoid a diverging
energy penalty, and so the layer thickness tends to zero. However,
physically there must be a transition to a constant layer thickness
as the angle $\Theta$ increases, i.e., there must be forces
preventing $\Theta\rightarrow \pi/2$ independent of the
layer-compression term.

The contribution $f_\mathrm{tilt} = f_\mathrm{tilt} (\sin \Theta)$
to the free energy density penalizes the deviation of the director
from the layer normal. To ensure a finite layer thickness, see the
discussion above, the general form of the tilt contribution will be
such that $f_\mathrm{tilt} (\sin \Theta) \to \infty$ as
$\Theta\rightarrow \pi/2$. When expanded to leading order in $\sin
\Theta$, it reduces to
\be
\label{fTilt}
f_\mathrm{tilt} = \half \scp \sin^2 \Theta \, ,
\ee
where $a_t$ is a coefficient that vanishes as the $A$-$C$ transition
is approached. For simplicity, we will work in the following with
the simple phenomenological form~(\ref{fTilt}).

Typical values of the constants are $B \sim 10^7$Pa, $\scp \sim 10^5-
10^6$Pa, and $\mu \sim 10^5-10^6$Pa. In determining the main features
of the material properties only the ratios of these values are
important. Consequently we will denote
\bea
b = B/\mu\;\;\; {\rm  and} \;\;\; c =\scp/\mu.
\eea
Note that the limit $c\to \infty$ locks the director to the layer
normal. On the other hand, when $c$ is small as it would be near the
AC transition, deviations of $\Theta$ from zero can be large.
Dominating these energy scales is that for volume change. The bulk
modulus is, as for all rubbers, of the order of $10^9 {\rm Pa}$
which means that deformations are at constant volume, that is
$\det{\ten{\lambda}}=1$ rigidly.

Note that a semi-soft term could also be included of the form
\be
f_\mathrm{semi} = \half \alpha \mu \textrm{Tr}\left[
  \ten{\lambda}\cdot (\ten{\delta} - \vec{n}_0 \vec{n}_0) \cdot
  \ten{\lambda}^{T} \cdot \vec{n}\vec{n}\right] .
\ee
The value of $\alpha$ can be estimated from either the soft plateau in
elastomers, or from the electroclinic effect, and is typically found
to be $\alpha < 0.1$. It turns out that this term does not affect the
qualitative features of the mechanic response to the deformations that
we consider. We will discuss this issue briefly and exemplarily in our
analysis of $xz$-shear, but we will not scrutinize the effects of the
semi-soft term in detail.

\section{Example Geometries}

\begin{figure}[!htb]
  \includegraphics[width = 0.45\textwidth]{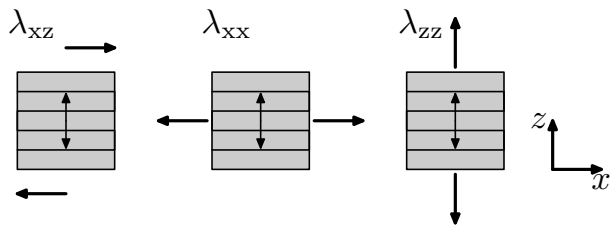}
\caption{The imposed shear, in-plane elongation and out of plane
elongation (and compression) geometries that will be considered in
  this paper. }
\label{fig:geometries}
\end{figure}

Three deformation geometries will be considered here, and  are shown
in Fig.~\ref{fig:geometries}, together with the initial director and
layer normal which are assumed to be along $\vec{z}$.

The model described above has a complicated form because
incompressibility forces the appearance of a cofactor of the
deformation gradient $\ten{\lambda}$. It is possible to eliminate
the cofactor  dependence by recognizing that the director and layer
normal move in a $2D$ subspace, which we take to be the $xz$ plane
(where $\vec{z}$ is parallel to the initial layer normal). Then all
the actual calculations can be made using a $2D$ representation of
this model outlined in appendix \ref{app:2dmodel}.

\subsection{Imposed $\lda{xz}$}
\label{sec:imposedxz}

First we examine the imposed shear deformation as depicted in
Fig.~\ref{fig:geometries}. Here the following will be used
\be
\ten{\lambda} = \left(\begin{array}{ccc}
\lda{xx}&0&\lda{xz}\\
0&\lda{yy}&0\\
0&0&\lda{zz}
\end{array}\right)
\;\;
;
\;\;
\begin{array}{ccc}
\vec{n}_0 &=& (0, 0, 1)\\
\vec{n} &=& (\sin \theta, 0, \cos \theta).
\end{array}\label{eq:defgrad}
\ee
To ensure incompressible response, $\det{\ten{\lambda}}=1$, one takes
$\lda{yy} = 1/(\lda{xx}\lda{zz})$.  For this choice of $\ten{\lambda}$, the
new layer normal is $\ten{\lambda}^{-T} \cdot \vec{k}_0 =
\vec{k}_0/\lda{zz}$. Layers are unrotated, but generally dilated by
$\lda{zz}$. The deviation of $\vec{n}$ from $\vec{k}_0$, that is
$\Theta$, can thus be identified with the usual $\theta$, the rotation
of the director. We use $\theta$ until later when there is layer
rotation and the distinction must be drawn. On substituting these
expressions into the free energy density we obtain the following
expression
\bea
\nonumber
f &=& \half B \left(\frac{\lda{zz}}{\cos \theta} - 1\right)^2+ \half \scp \sin^2 \theta\\
\label{eqn:xzfed}
&+& \half \mu \left( \frac{1}{\lda{xx}^2 \lda{zz}^2}+ \lda{xx}^2 \left[\cos^2\theta+\frac{1}{r}\sin^2\theta\right]\right.\\
&+&(\lda{zz}\cos\theta + \lda{xz} \sin\theta)^2 + r (\lda{zz}\sin \theta -
\lda{xz} \cos \theta)^2\bigg) .\nonumber
\eea
For this $\ten{\lambda}$, the additional semi-soft term in the free
energy is $f_\mathrm{semi} = \half \alpha \mu \lda{xx}^2 \sin^2
\theta$. It does not change the qualitative behavior of the
elastomer, so will not be considered here.

This free energy should now be minimized with respect to $\theta$,
$\lda{xx}$ and $\lda{zz}$. It is straight forward to find the minimal
value of $\lda{xx}$:
\be
\lda{xx}^4 = \frac{1}{\lda{zz}^2 (\cos^2 \theta + \frac{1}{r} \sin^2 \theta)}.
\ee
We now assume that $B$ is much larger than $\scp$ and  $\mu$ so that
we can approximate $\lda{zz} \approx \cos \theta$. Such an
identification is forced by the first term of Eq.~(\ref{eqn:xzfed})
when its coefficient is large. The leading order response for
$\lda{xz}\ll 1$ is then (recalling that $c$ is the reduced
angular modulus $a_t/\mu$):
\bea
\theta &=& \frac{(r-1)r \lda{xz}}{cr + (r-1)^2} ,\\
f&=&\frac{3\mu}{2}+ \frac{\mu}{2}\frac{c r^2 \lda{xz}^2}{cr + (r-1)^2} , \\
\sigma&=& \mu\frac{c r^2 \lda{xz}}{cr + (r-1)^2}\label{eq:stress1},
\eea
where $\sigma$ is the nominal or engineering stress
$\partial f/\partial \lda{xz}$. Asymptotic analysis can also be
performed for large deformations, but it should be remarked that
whilst the assumptions of very large strains seem unrealistic, the
results obtained from asymptotic methods are usually applicable to a
much larger region than anticipated. Here we assume that $\lda{xz}\gg
1$, and $\lda{zz}\ll 1$, and again we have to leading order:
\bea
\lda{zz} &=& \frac{1}{(r-1)^{1/3}r^{1/6}\lda{xz}^{2/3}},\\
f&=&\frac{\mu}{2}\lda{xz}^2\, , \quad \sigma=\mu \lda{xz}.
\eea
Shear across an unmoving director has a modulus $r\mu$, as for
instance inspection of Eq.~(\ref{eq:stress1}) in the $c \gg 1$ limit
reveals. Here in this unphysical limit, the modulus has dropped to
$\mu$, indicative of minimal chain extension in the gradient direction
of the shear.  Thus the form of the stress indicates that the director
rotation is very large, evidently $\sim \pi/2$. As discussed above,
this extreme director rotation is an artifact of the reduction of
$f_{\text{tilt}}$, which in its full form has to suppress the approach
of $\Theta$ to $\pi/2$ on physical grounds.

\begin{figure}[!htb]
  \includegraphics[width = 0.48\textwidth]{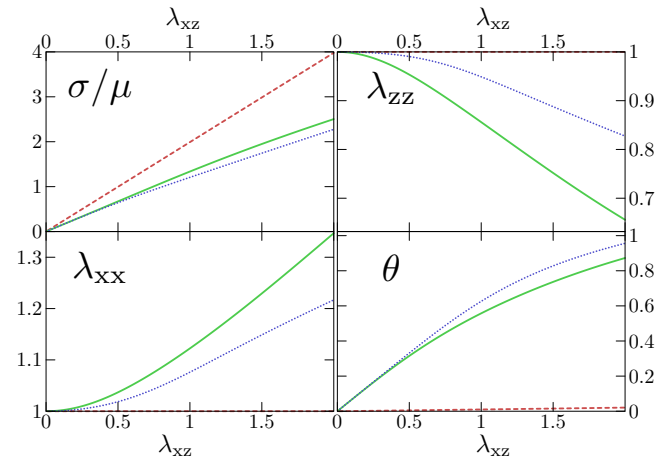}
\caption{The nominal shear stress, the deformation components $\lda{zz}$
  and $\lda{xx}$, and the rotation angle $\theta$ of the director for an
  imposed $\lda{xz}$ deformation. The dashed (red) line has $(b,c,r) =
  (60, 100, 2)$, the solid (green) line has $(60, 1,2)$ and the dotted
  (blue) line has $(1, 1, 2)$.}
\label{fig:xzcomponents}
\end{figure}

\begin{figure}[!htb]
  \includegraphics[width = 0.48\textwidth]{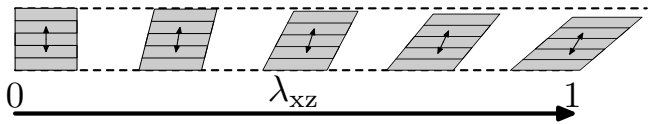}
\caption{An illustration of the rotation of the director, and the
  sympathetic shears for an imposed $\lda{xz}$ deformation, with $(b, c,
  r) = (60, 1, 2)$ as on the solid (green) line in
  Fig.~\ref{fig:xzcomponents}.  }
\label{fig:xzcartoon}
\end{figure}

To illustrate the intermediate features of the model,
Fig.~\ref{fig:xzcomponents} shows the numerical solution to the
minimization problem, and Fig.~\ref{fig:xzcartoon} shows an
illustration of the deformation of the elastomer.
Since the $xx$ and $zz$ response is independent of the sign of
$\lda{xz}$, on symmetry grounds for small imposed shear, $\lda{xx} -1 \sim
\lda{xz}^2$ and $\lda{zz} \sim - \lda{xz}^2$. Rotation does sense the sign
and hence $\theta \sim \lda{xz}$. For small values of $c$, $\theta$
saturates at $\pi/2$ for large deformation. By contrast, for large
values of $c$ the director rotation is suppressed, in agreement with
\cite{adams:05a}.

In the shear experiments of Kramer and Finkelmann~
\cite{kramer-2007, KraFin2008} the applied deformations are similar
to the one that we just discussed. There is, however, the difference
that these experiments use setups (tilter and slider) that pre-set
$\lda{zz}$ as shear proceeds, and hence free relaxation of $\lda{zz}$ as
above is not possible. A detailed analysis of these experiments is
given in \cite{stenull07}.

\subsection{Imposed $\lda{xx}$}
\label{sec:imposedxx}
\begin{figure}[!htb]
  \includegraphics[width = 0.48\textwidth]{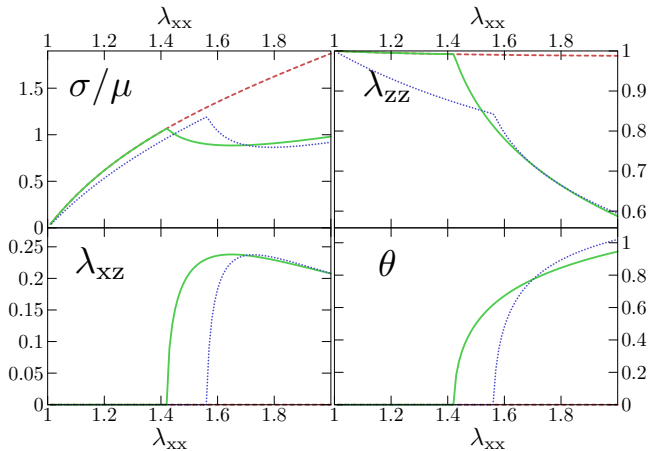}
\caption{The nominal shear stress, rotation angle, and deformation
  tensor components for an imposed $\lda{xx}$ deformation. The key is
  the same as in Fig.~\ref{fig:xzcomponents}.  }
\label{fig:xxcomponents}
\end{figure}

The deformation tensor is still that of Eq.~(\ref{eq:defgrad}), and
the free energy takes the same form Eq.~(\ref{eqn:xzfed}) as in the
previous section. The relaxation behavior of the system is sketched
in Fig.~\ref{fig:xxcomponents}, which illustrates several
interesting features including a threshold at which rotation of the
director starts, and a non-monotonic stress-strain curve.
\begin{figure}[!htb]
  \includegraphics[width = 0.48\textwidth]{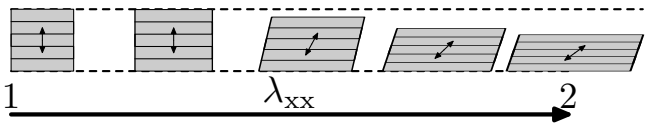}
\caption{An illustration of the rotation of the director, and the
  sympathetic shears for an imposed $\lda{xx}$ deformation,
  corresponding to the solid (green) line in
  Fig.~\ref{fig:xxcomponents}. }
\label{fig:xxcartoon}
\end{figure}

Analytically it is possible to obtain an expression for the
threshold value of $\lda{xx}$ at which the instability starts.
Minimization of the free energy in Eq.~(\ref{eqn:xzfed}) with
respect to $\lda{xz}$ results in the following expression
\be
\lda{xz} = \frac{(r-1) \lda{zz} \cos \theta \sin \theta}
{\sin^2 \theta
  + r \cos ^2 \theta} \, .
\ee
We again consider the case where $B\gg \scp, \mu$, and so $\lda{zz}
\approx \cos \theta$. This leaves only the variable $\lda{zz}$ to be
minimized over in the problem. For large $B$ the threshold occurs when
there is a minimum at $\lda{zz} = 1$, which results in the following
condition on $\lda{c}$, the critical value of $\lda{xx}$,
\be
\lda{c}^4 (r-1) - \lda{c}^2 (c r-1) -r =0.
\ee
Consequently, the threshold is given by
\be
\lambda_{c}^2 = \frac{(cr -1) + \sqrt{4 (r-1)r + (cr -1)^2}}{2 (r-1)}.
\ee
This threshold is unphysically large for $c\gg 1$ and so it would be
inaccessible to mechanical experiments in this limit. But when $r=2$
and $c=1$, we have $\lda{c} = \sqrt{2}$, in agreement with the
numerical results presented in Fig.~\ref{fig:xxcomponents} and
possibly within the range of experiments.

The behavior of the system for both small and large values of the
deformation can again be analyzed. Before the threshold, we have
$\theta = \lda{xz} = 0$, and $\lda{zz} = 1$. Consequently
\bea
f &=& \half \mu \left(1+\lda{xx}^2 + \frac{1}{\lda{xx}^2} \right) , \\
\sigma &=&\mu \left(\lda{xx} - \frac{1}{\lda{xx}^3}\right) ,
\eea
which is the $2D$ rubber elastic response seen experimentally
\cite{nishikawa1999} in the case where the director and layer normal
are rigidly anchored, and described theoretically in this framework
in \cite{adams:05a}. After the threshold we have for $\lda{xx}\gg1$
\bea
\lda{zz} &\approx& \left(\frac{r}{r-1}\right)^{1/4}\frac{1}{\lda{xx}} \, ,\\
\theta &\approx& \pi/2-\left(\frac{r}{r-1}\right)^{1/4}\frac{1}{\lda{xx}} \, ,\\
\lda{xz}&\approx&\frac{\sqrt{(r-1)r}}{\lda{xx}^2} \, ,\\
f&\approx&\frac{\mu}{2 r} \lda{xx}^2 \, .
\eea
This limit is again extreme and unphysical, but useful for
understanding trends.
 The result is the same as that
for a nematic rubber that is being stretched perpendicular to its
director, the director subsequently rotating to be parallel to the
applied elongation, the $x$-direction. The $1/r$ reduction in the
effective modulus arises because there is an effective
$xx$-elongation of $\sqrt{r}$ on director rotation and the effective
extension with respect to this state is only $\lda{xx}/\sqrt{r}$.

The induced shear at the threshold $\lda{xx} = \lda{c}$ increases
infinitely quickly with stretch,  as is the case in theory and
experiment \cite{Kundler1a} for the response at the threshold when
nematic elastomers are stretched perpendicular to their initial
director. A symmetry argument shows that this must be the case: the
instability is insensitive to the signs of the angle $\theta$ and
the shear $\lda{xz}$, and the stretch $\lda{xx}$ does not distinguishing
among these signs. Thus one must have $\lda{xz}^2 \sim \lda{xx}-\lda{c}$
and $\theta^2 \sim \lda{xx}-\lda{c}$ (or higher even powers of $\lda{xz}$
and $\theta$). On taking roots, one has $\lda{xz} \sim \pm
\sqrt{\lda{xx}-\lda{c}}$ and $\theta \sim \pm \sqrt{\lda{xx}-\lda{c}}$, that
is, singular growth at $\lda{c}$.

The results here clearly show a non-monotonic stress-strain relation
that is not seen, for instance, in theory or experiment for Sm-$A$
elastomers at their instability under a $\lda{zz}$ extension along
their layer normal (see subsection~\ref{sect:imposedzz} below). In
the $zz$ extension case, the layer normal rotates away from the
extension direction to allow in-plane stretch and shear. But in
rotating away, the layer normal takes the
nematic director with it and thus towards the contraction
diagonal associated with the shear that is growing in a singular
fashion. For prolate chains, this compression along the
director of the naturally long dimension of the chain distribution
costs additional energy.  This is still a lower energy path than
that associated with layer extension which would be suffered if
layers did not rotate. By contrast here in the case of in-plane
stretch, Fig.~\ref{fig:xxcartoon} shows that when director rotation
takes place, it is instead towards the direction of the
\textit{extension} diagonal of the induced $xz$-shear and thus the
naturally longer dimension of the chain distribution is more
accommodated. This shear grows in a singular fashion. The consequent
slow down in the growth of the elastic free energy is sufficient to
make the slope of the stress-strain relation negative.

Just as for the van der Waals gas, this type of non-monotonic
stress-strain curve is mechanically unstable to disproportionation.
It can be rectified by taking a mixture of the rotated and unrotated
states, and making a Maxwell construction on the stress-strain
relation. Experimentally, it is expected that there will be a
plateau in mechanical behavior, that is reversible. This is highly
unusual elastic behavior for a solid.

\begin{figure}[!htb]
  \includegraphics[width = 0.48\textwidth]{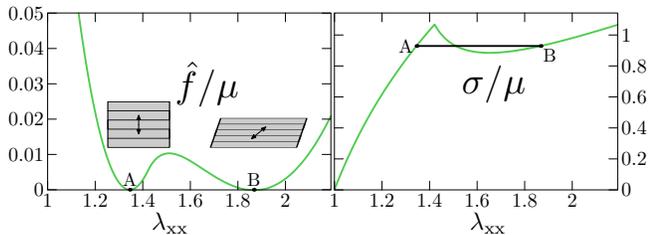}
\caption{An illustration of the free energy and stress for the non-monotonic
  stress-strain. The curve illustrated has $(b, c, r) = (60, 1, 2)$. Here
  $\lda{A} = 1.34$, $\lda{B} = 1.87$, and $\sigma_p = 0.978 \mu$. }
\label{fig:NMxxwells}
\end{figure}

\begin{figure}[!htb]
  \includegraphics[width = 0.48\textwidth]{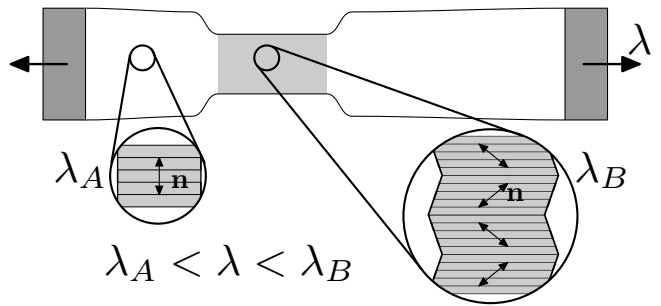}
\caption{An illustration of a possible microstructure that would
  result from the applied in-plane deformation. Depending on the cost
  of the interface the disproportionation illustrated here may happen
  in several regions in the sample. The region where the
  microstructure forms may be opaque as illustrated. }
\label{fig:NMxxmicro}
\end{figure}

To carry out the Maxwell construction, we define $\hat{f} = f -
(\sigma_p \lambda_{xx} + f_0)$, i.e., we subtract from he free
energy the common tangent that touches $f(\lda{xx})$ at the points of
coexisting strain with the same engineering stress $\sigma_p$, and
we determine $\sigma_p$ and $f_0$ such that $\hat{f}$ just touches
$\hat{f}=0$ at two points. An illustration of $\hat{f}$ which
highlights its non-convexity is shown in Fig.~\ref{fig:NMxxwells},
together with an illustration of the deformation gradient at the two
minima. The stress is also shown, with a plateau, that is formed by
mixing the two deformation gradients illustrated. The anticipated
microstructure may be as illustrated in Fig.~\ref{fig:NMxxmicro},
however the effects of surface energy and interfacial regions may
lead to a different structure.
The actual region of the singular shear and director
rotation seen in Fig.~\ref{fig:xxcomponents} is thus jumped across
and these interesting responses may not be observable.

The value of the threshold provides a useful test for the value of
the modulus $\scp$. As no threshold has been observed through
in-plane elongation experiments, it is thought that $\scp$ should be
comparable to or larger than $\mu$. For some samples, these
stretching experiments have been performed up to a strain of
$\lda{xx}=2.5$, but no deviation from linearity has been observed
\cite{kramer07}. This suggest that the anchoring is so strong in
some elastomers, that the sample ruptures before the director and
the layer normal become unlocked.

\subsection{Imposed $\lda{zz}$}\label{sect:imposedzz}

It is well-known that in liquid Sm-$A$ \cite{Clark:73} and in
elastomeric Sm-$A$ \cite{nishikawa1999} there is an instability when
stretching parallel to the layer normal. For small deformations the
elastomer simply elongates, however for large deformations above a
threshold, layers start to rotate, and the sample undergoes
effective in-plane shear, as first predicted for
elastomers within continuum elasticity \cite{Terentjev:94a} and
previously analyzed in the framework of rubber elasticity
\cite{adams:05a} but with $\vec{n}$ and $\vec{k}$ rigidly locked ($c
= \infty$). When $c$ is finite, elongation parallel to the layer
normal still results in the same instability, as is already known
from Lagrangian elasticity methods \cite{stenull:011706}. More
unusually, as a consequence of finite $c$, we predict
that compression ($\lda{zz}<1$) can also result in a rotational
instability, but only of the director with an unrotating layer
system.

\subsubsection{Elongation $(\lda{zz}>1)$}
Contrary to the deformations considered thus far, elongation along the
initial layer normal generates layer rotation, and we are compelled to
distinguish between director and layer rotation. Thus we take $\Theta$
to denote the angle of the director with respect to the current layer
normal, and we introduce $\zeta$ as the angle by which the layers
rotate relative to the initial layer normal. The deformation gradient
tensor
\be
\ten{\lambda} = \left( \begin{array}{ccc}
\lda{xx} & 0&0\\
0&\frac{1}{\lda{xx}\lda{zz}} & 0\\
\lda{zx} & 0 & \lda{zz}\end{array} \right) \label{eq:firstshear}\ee
must be assumed since $\lda{zx}$ shears are those that produce layer
rotations, see Eq.~(\ref{eq:layers}). This awkward form of the
deformation gradient can be decomposed to reveal the true
deformations by incorporating a rotation matrix. We first deform the
system by $\ten{\Lambda}$ which we take to be as in the previous
cases, i.e. as in Eq.~(\ref{eq:defgrad}) with stretches and instead
an $xz$-shear. The layer normal, given this type of deformation, is
as yet unrotated. We then perform a body rotation on both the
elastomer and the director, which of course leaves the free energy
invariant. The rotation can be chosen to transform the $xz$ shear of
$\ten{\Lambda}$ into the $zx$ shear of $\ten{\lambda}$ in
(\ref{eq:firstshear}). The overall deformation is made up as follows
\be \ten{\lambda} = \ten{R}(\zeta) \cdot \ten{\Lambda} \; ; \;\;\;
\ten{\Lambda} = \left( \begin{array}{ccc}
    \CLDA{xx} & 0&\CLDA{xz}\\
    0&\CLDA{yy} & 0\\
    0 & 0 & \CLDA{zz}\end{array} \right),\label{eq:composeddefs}
\ee
where $\ten{R}(\zeta)$ represents a rotation through an angle
$\zeta$ about $y$. At this point we can see that all
mechanical  deformations of Sm-$A$ can be reduced to the three of
Fig.~\ref{fig:geometries} -- others can be reduced to these by
suitable rotations.

\begin{figure}[!tb]
  \includegraphics[width = 0.48\textwidth]{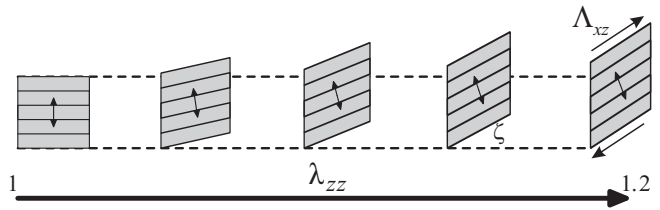}
\caption{An illustration of the rotation of the director, and the
sympathetic shears for an imposed $\lda{zz}$ elongation.  The $\CLDA{xz}$
shear is shown in the frame of the layer system.  $\zeta$ is the
shear angle and the rotation angle to go from the lab to the layer
frames.  } \label{fig:zzcartoon}
\end{figure}

Fig.~\ref{fig:zzcartoon} shows the development of the $\lda{ij}$,
including $\lda{zx}$ in response to $\lda{zz}$. The last sketch has the
$\CLDA{xz}$ marked in as it would appear in the frame of the body. The
rotation angle $\zeta$ can be determined by demanding that the
$\CLDA{zx}$ component is zero.  Practically, this is done by putting in
the explicit form for $\ten{R}$ in the above, evaluating
$\ten{\Lambda} = \ten{R}^T\cdot \ten{\lambda}$ and inspecting
$\CLDA{zx}$. It is also clear geometrically from the second to last
frame of Fig.~\ref{fig:zzcartoon} that to convert $\ten{\Lambda}$ to
the lab frame $\ten{\lambda}$ one needs to rotate by the shear angle
$\zeta = \tan^{-1}(\CLDA{xz}/\CLDA{zz})$, resulting in $\cos \zeta =
\frac{\CLDA{zz}}{\sqrt{\CLDA{xz}^2 + \CLDA{zz}^2}}$. The connection between
the new deformation components $\CLDA{ij}$ and the old $\lda{ij}$ is then
\bea
\lda{xx} &=& \frac{\CLDA{xx}\CLDA{zz}}{\sqrt{\CLDA{xz}^2 + \CLDA{zz}^2}} \, ,\\
\lda{zx} &=& \frac{\CLDA{xx}\CLDA{xz}}{\sqrt{\CLDA{xz}^2 + \CLDA{zz}^2}} \, ,\\
\lda{zz} &=& \sqrt{\CLDA{xz}^2 + \CLDA{zz}^2} \, , \label{eq:newzz}
\eea
see
\cite{adams:05a} for explicit discussion.

\begin{figure}[!tb]
  \includegraphics[width = 0.48\textwidth]{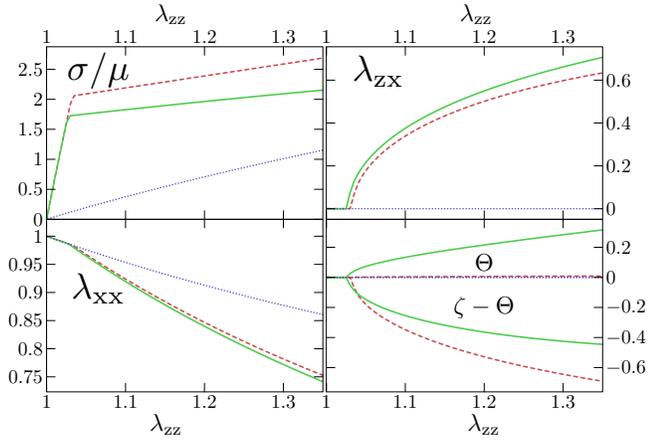}
\caption{The nominal shear stress, rotation angle, and deformation
  tensor components for an imposed $\lda{zz} >1$ deformation. The key is as
in Fig.~\ref{fig:xzcomponents}. Note that the dashed (red) curve for
$\Theta$ lies above zero for $\lda{zz}$ exceeding the threshold,
although this is hard to see because $\Theta$ is very small.}
\label{fig:zzcomponents}
\end{figure}

Recall $\Theta$ is the angle between the layer normal and the
director. Since the body and the system of layers is rotated by
$-\zeta$, the director can be expressed as
\be
\vec{n} = \left(\sin [\Theta-\zeta], 0, \cos[\Theta - \zeta]\right),
\ee
so that after rotation by $\zeta$ in (\ref{eq:composeddefs}) it
becomes \mbox{$\vec{n} = \left(\sin [\Theta], 0,
    \cos[\Theta]\right)$}.  After this transformation is performed, we
obtain the free energy density again given by Eq.~(\ref{eqn:xzfed})
with $\lda{ij} \rightarrow \CLDA{ij}$ and $\theta\rightarrow \Theta$.
However, we do not impose $\CLDA{zz}$, but still impose $\lda{zz}$,
whereupon $\CLDA{zz} = \sqrt{\lda{zz}^2-\CLDA{xz}^2 }$ on rearranging
(\ref{eq:newzz}). The behavior of the system can be analyzed
numerically and results for this analysis are presented in
Fig.~\ref{fig:zzcomponents}.
Note we still see a threshold behavior, but now in addition a
rotation of the director with respect to the layer normal $\Theta$.
From the decomposition discussed above, it is clear the
  rotation of the director with respect to the layer normal arises
  because of the effective $xz$ shear. Consequently we expect a
  non-zero value of $\Theta$ on symmetry grounds as observed in \S
  \ref{sec:imposedxz}.

There are several ways that the position of the threshold can be
computed. One method is to expand the free energy density for small
values of $\Theta$. We can then minimize with respect to $\Theta$,
and substitute in the minimal value. It is then possible to minimize
over $\CLDA{xx}$, and substitute this back the free energy. Only
$\CLDA{xz}$ remains to be minimized over. Setting the second derivative
of this expression with respect to $\CLDA{xz}$ to zero at $\CLDA{xz} = 0$
then gives the equation for the threshold, because it gives the
point where the level sets change from being convex to concave. This
results in the following polynomial for the critical value $\lda{c}$
of $\lda{zz}$
\bea
\nonumber
&&b^2 r(\lda{c}-1)^2\lda{c}^4 - b(\lda{c}-1)\lda{c}^2(2 r -1) - (r-1) (\lda{c}^3 -1)\\
&&+cr \lda{c}\left[(b-r+1)\lda{c}^3-b \lda{c}^2 -1 \right]=0.
\label{eqn:zzthreshold} \eea
The terms in $c$ have been collected together here, so that it is
clear that for $c\gg 1$ the same polynomial in square brackets as in
the locked director case \cite{adams:05a} is recovered. In
that case there was only an instability if the reduced layer spacing
modulus was large enough, $b > r-1$.  Now if the director is no
longer locked to the layer normal, finite $c$, then there is
naturally an instability for even smaller values of $b$ since the
lower energy route of rotation is more accessible if the director
can rotate towards the extension diagonal associated with the
concomitant shear.

In the limit $B\gg \scp, \mu$ for which $\lda{c}$ exists,
\be
\lda{c}\approx
1+ \frac{2 r-c r-1+\sqrt{(c r+1) \left(4 r^2+c r-4  r+1\right)}}{2 b r} .
\ee
For $r=2$ and $b=60$ we have:
\bea
\lim_{c\rightarrow\infty}\lda{c} &\approx& 1+\frac{r}{b} \approx 1.033 \, ,\\
\lim_{c\rightarrow1}\lda{c} &\approx&1.028 \, , \eea
where the limits have been taken from Eq.~(\ref{eqn:zzthreshold}).
Consequently the experimental evaluation of this threshold is not
sensitive to the value of $c$, and does not discriminate between the
weak anchoring, and the locked layer normal theories.

For small deformations (before the threshold is reached) we have
\bea
\lda{zx} &=& \Theta = 0 \, ,\\
\lda{xx} &=& \frac{1}{\lda{zz}} \, ,\\
f&=&\half B (\lda{zz}-1)^2 +\half \mu\left( \frac{2}{\lda{zz}} + \lda{zz}^2\right) ,
\label{eqn:smallzz}
\eea
which is identical to the locked layer normal case. 

Before moving on to compression, we find it worthwhile to compare
our results  to the available experimental findings and other
theoretical predictions. The stress-deformation curves for $B \gg
\mu$ are in absolute agreement with the experimental curves by
Nishikawa and Finkelmann~\cite{nishikawa1997} as well as the
theoretical predictions of Refs.~\cite{adams:05a} and
\cite{stenull:011706}. The results, or at least their
interpretation, differ, however, as far as $\Theta$ is concerned. As
mentioned above, the rubber-elastic model of Ref.~\cite{adams:05a}
assumes that the layer normal and the director are rigidly locked
and thus inevitably produces $\Theta=0$ for any value of $\lda{zz}$.
From their X-ray data, Nishikawa and Finkelmann conclude that there
is no relative tilt between layer normal and the director below and
above the threshold. The Lagrangian model of
Ref.~\cite{stenull:011706} predicts nonzero but small $\Theta$ above
the threshold, like in the dashed (red) curve in
Fig.~\ref{fig:zzcomponents}. Given that the angle-resolution in the
experiment of Nishikawa and Finkelmann was of the order of one
degree, it is possible that there was Sm-$C$ like tilt in the
experiment that was too small to detect if $\Theta$ followed a curve
similar to the dashed (red) curve in Fig.~\ref{fig:zzcomponents}. In
this case, there is no contradiction between the experimental data
and the theoretical findings of Ref. \cite{stenull:011706} and the
present paper.

\subsubsection{Compression $(\lda{zz}<1)$}

\begin{figure}[!htb]
  \includegraphics[width = 0.48\textwidth]{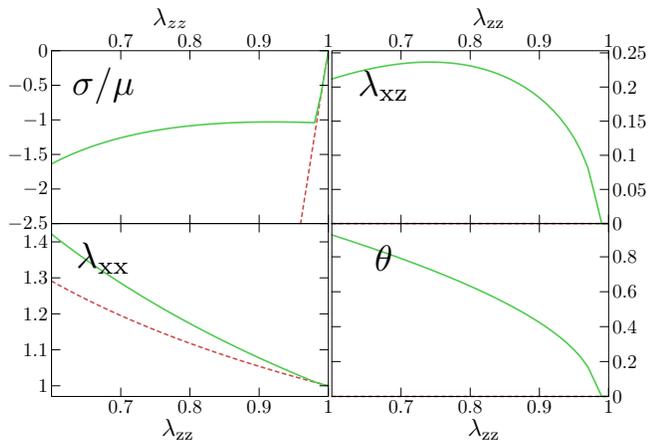}
\caption{The nominal shear stress, rotation angle, and deformation
  tensor components for an imposed $\lda{zz}$ deformation. The key is as
  in Fig.~\ref{fig:xzcomponents}. Note that the stress-deformation curve features a slight negative slope that is hard to see.}
\label{fig:zzccomponents}
\end{figure}

\begin{figure}[!htb]
\includegraphics[width=0.48\textwidth]{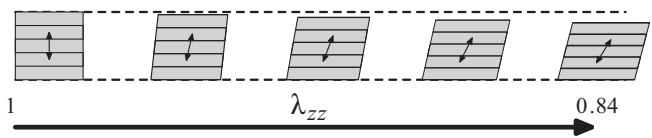}
\caption{An illustration of the rotation of the director, and the
  sympathetic shears for an imposed $\lda{zz} <1$ compression.}
\label{fig:zzccartoon}
\end{figure}

For the compression case we again use the free energy expression of
Eq.~(\ref{eqn:xzfed}) since in the absence of a $\lda{zx}$ there is no
layer rotation. Compression along the layer normal is resisted by the
layer spacing potential. There is also a nematic rubber elastic
penalty for the chain compression. Nematic elastomers in theory and
experiment \cite{Verduzco:147802} are known to reduce their elastic
energy on compression along the director by rotation of the director,
thereby presenting a shorter dimension of their chain distribution to
the imposed strain. We find an analogous effect here. Numerical
results for the compression case are shown in
Fig.~\ref{fig:zzccomponents}, and illustrated in
Fig.~\ref{fig:zzccartoon}.

There is again a threshold, $\lda{c}$, at which rotation starts in the
weak anchoring model. The threshold obeys the following polynomial:
\be
c r \lda{c} + b r \lda{c}^2(\lda{c}-1) + (r-1) (\lda{c}^3 -1) =0.
\ee
There is always a solution to this equation, that is, in
  principle an instability against compression should always exist.
For $B\gg \scp$, the solution is 
\begin{align}
\label{compressionThreholdLageB}
\lda{c} \approx 1-\scp/B \,  ,
\end{align}
which provides an important test for the magnitude of the parameter
$\scp$.  For example if $\scp\sim \mu$ and $B \sim 60 \mu$ then $\lda{c}
\approx 0.983$, i.e. a compression of a few percent. For large $c$ the
instability moves to (unphysically) large compressions, $\lda{c} \sim
0$. At small deformations, before any rotational instability, the
behavior is exactly as in the elongation case, given in
Eq.~(\ref{eqn:smallzz}).

The stress-deformation curve shown in Fig.~\ref{fig:zzccomponents}
is again non-monotonic, so is unstable in the region of negative
slope. This unphysical feature can be resolved by resorting to a
Maxwell construction as explained in Sec. \ref{sec:imposedxx}. As in
Sec.~\ref{sec:imposedxx}, this construction leads to the prediction
of a plateau in the stress-deformation curve and microstructure that
features, unless boundary conditions prevent this, a mixture of
sheared and compressed regions.

Reference~\cite{stenull:011706} focused on stretching along the layer
normal as in the experiments of Nishikawa and Finkelmann, and its
authors chose not to consider compression along the layer normal.
However, from the equations of Sec. III of Ref.~\cite{stenull:011706},
it is straightforward to see that the Lagrangian model also predicts
an instability towards shear for compression along the layer, $u_{zz}
<0$, where $\ten{u} = \frac{1}{2} (\ten{\lambda}^T \ten{\lambda} -
\ten{\delta})$ is the usual strain tensor. The thresholds $u_{zz}^c$
for the onset of $u_{xz}$ shear are determined by the values of
$u_{zz}$ for which the effective modulus $r_R (u_{zz})$ in the tilt
energy density passes through zero from positive to negative. For the
precise definition of the effective modulus, which is related to the
parameter $a_t$ defined via Eq.~(\ref{fTilt}), we refer to Eq.~(3.10b)
of Ref.~\cite{stenull:011706}. The equation $r_R (u_{zz}) = 0$ has 2
solutions, the one given in Eq. (3.11) of Ref.~\cite{stenull:011706},
and a corresponding one with the minus in front of the square root
replaced by a plus. The solution with the plus pertains to compression
and was, therefore, not discussed in Ref.~\cite{stenull:011706}. Based
on the available experimental data for fluid~\cite{ArcherDie2005} and
elastomeric smectics~\cite{brehmer&Co_1996}, the Lagrangian theory of
Ref.~\cite{stenull:011706} produces the estimates $u_{zz}^c \approx -
0.025$ and $u_{zz}^c \approx -0.0025$, respectively, which are
consistent with estimates for $\lambda_c$ based
Eq.~(\ref{compressionThreholdLageB}) and the typical values for $\scp$
and $B$ quoted in Sec.~\ref{modelFreeEnergy}.

\section{Discussion and Conclusion}

We have explored the consequences of having a weakly anchored director
in a microscopic model of a Sm-$A$ elastomer for three deformations:
in-plane shear, in-plane elongation, and deformation parallel to the
layer normal. For the in-plane shear, it is found that the director
rotates toward the extension diagonal and that initially the rotation
angle is proportional to the amplitude of the shear applied. For
in-plane elongation, and for compression and elongation parallel to
the layer normal, the possibility of rotation of the director leads to
the prediction of instabilities of the system to director rotation.

The stress-deformation curves predicted by our model for in-plane
elongation and for compression parallel to the layer normal are
non-monotonic.  Since a region of negative slope in the
stress-deformation curve is unstable, this behavior is unphysical and
will not be seen experimentally. Instead the elastomer should form a
mixture of two different deformations, and exhibit a plateau in the
stress-deformation curve, as illustrated in \S \ref{sec:imposedxx}. In
the case of imposed in-plane elongation disproportionation has not
been reported experimentally; it may, however, be possible to engineer
elastomers in which this behavior could be observed.

The instability toward the development of shear for elongation along
the layer normal has been discussed theoretically in earlier
papers~\cite{adams:05a,stenull:011706}. Here, we also predict an
instability to rotation of the director under compression along the
layer normal. This could be analyzed experimentally using the samples
in which the director and layer normal are parallel to the normal to
the film as reported in \cite{nishikawa04}. The thresholds at which
the different instabilities occur provide a useful way to determine
the model parameters experimentally, and to find the value of the
parameter $\scp$.

\begin{acknowledgments}
  We are grateful for support from the EPSRC (JMA and MW) and the
  National Science Foundation under grants DMR 0404670 and MRSEC
  DMR-0520020 (TCL and OS). We thank Dominic Kramer for discussing
  in-plane extension experiments of smectic elastomers.
\end{acknowledgments}

\appendix*

\section{General $2D$ model formulation}
\label{app:2dmodel}
 We give here a practical method of
calculating the free energy in the complex situations of extensions,
shears and rotation. Simplification is possible because the
deformations occur in the $xz$ subspace of $\ten{\lambda}$ and are
denoted by
\be
\ten{G} = \left( \begin{array}{cc} \lda{xx} & \lda{xz}\\
\lda{zx} & \lda{zz}\end{array}\right), \ee
with a single $y$ deformation ($\lda{yy} = \det{\ten{G}}^{-1}$)
that acts to preserve volume.  Using this notation, and assuming
that the director, $\vec{n}$ remains in the $xz$ plane, then the
nematic free energy density is
\be
f_{el}=\half \mu \left\{ \textrm{Tr}\left[ \ten{G} \cdot \ten{\ell}_0
    \cdot \ten{G}^{T} \cdot \ten{\ell}^{-1}\right] +
 (\textrm{det}\ten{G})^{-2}\right\}.
\ee
The layer normal can be calculated \cite{adams:05a} as follows
\bea
\vec{k}\propto\ten{\lambda}^{-T} \cdot\vec{z} &=&
(\ten{\lambda} \cdot \vec{x}) (\times \ten{\lambda} \cdot \vec{y})=\frac{(\ten{G}\cdot\vec{x})\times \vec{y} }{\textrm{det}\ten{G}} \, ,\\
|\ten{\lambda}^{-T} \cdot\vec{z}|&=&\frac{|\ten{G} \cdot \vec{x}|}{\textrm{det}\ten{G}} \, .
\eea
The last term we require is the dot product of $\vec{n}$ and $\vec{k}$
to calculate the angle between the layer normal and director.
\bea \nonumber \vec{n}\cdot \vec{k} \!&=&\! \frac{((\ten{G} \cdot
\vec{x}) \times \vec{y})\cdot \vec{n}}{\ten{G}\cdot \vec{x} \times
\vec{y}}
=\frac{\sqrt{(\vec{n} \times (\ten{G}\cdot \vec{x})) \cdot(\vec{n} \times (\ten{G} \cdot \vec{x}))}}{|\ten{G}\cdot\vec{x}|} \\
&=&\left[1-\left( \vec{n} \cdot
\widehat{\ten{G}\vec{x}}\right)^2\right]^{1/2}   ,
\eea
where $\widehat{\ten{G}\vec{x}}$ denotes the unit vector $\ten{G}
\vec{x}/|\ten{G} \vec{x}|$. Combining the above results produces the
following total free energy density expression
\bea
\nonumber
f &=& \half \mu \left\{\textrm{Tr} \left[ \ten{G} \cdot \ten{\ell}_0 \cdot \ten{G}^{T} \cdot \ten{\ell}^{-1}\right] + \frac{1}{(\textrm{det} \ten{G})^2}\right\}\\
\nonumber
&+&\half B \left( \textrm{det}\ten{G}/\left[|\ten{G}\cdot\vec{x}|^2 - (\vec{n}\cdot \ten{G} \cdot \vec{x})^2\right]^{1/2}-1\right)^2\\
&+&\half \scp\left( \vec{n} \cdot  \widehat{\ten{G}\vec{x}} \right)^2 .
\label{eqn:fed}
\eea
This expression no longer involves the cofactor of the deformation
gradient.

\bibliographystyle{apsrev}  


\end{document}